\documentclass{article}

\usepackage{epsfig}
\usepackage[english]{babel} 
\usepackage{amsmath}
\usepackage{amsfonts}

\title{Interaction Strengths for the Fock-Space Formulation
       of the Nuclear Pairing Problem\thanks{This work has been partly
       supported by the Polish Committee for Scientific Research under 
       Contract No.~2P~03B~115~19 and the collaboration between IN2P3 and 
       Polish Laboratories nr. 99-95}}

\author{ J.~Dudek$^1$, \underline{K.~Mazurek}$^{1,2}$ and B. Nerlo-Pomorska$^2$
\\ 
{\it $^1$Institut de Recherches Subatomiques,
              IN$_2$P$_3$-CNRS/Universit\'e Louis Pasteur} \\
          {\it  F-67037 Strasbourg Cedex 2, France}\\
{\it $^2$Katedra Fizyki Teoretycznej, 
              Uniwersytet Marii Curie-Sk\l{}odowskiej,}\\
              {\it PL-20031 Lublin, Poland}
       }
 
\begin{document}

\maketitle

\begin{abstract}
A realistic nuclear mean-field hamiltonian with pairing has been diagonalized
using Fock space representation that allows for nearly exact treatment of the
problem. Calculations were performed for all the even-even nuclei with 
$Z \in (20, 100)$, whose pairing gaps were possible to extract out of the
experimental masses. The optimal values of the pairing strength constants for
the protons and neutrons have been found.
\end{abstract}


\vspace{1cm}

In the large scale microscopic calculations of the nuclear total-energy
surfaces the method of Strutinsky plays an important role allowing for fast and
fully automatic calculations related to the equilibrium deformations, shape
coexistence, fission probabilities and many other mechanisms and phenomena. In
the related formalism the calculation of shell- and pairing-energies plays a
decisive role, the latter obtained so far with the help of the Bogolyubov
transformation and the associated Bardeen-Cooper-Schrieffer approximate method.
The new method proposed in Ref.~\cite{mol} allows to obtain the exact (in some
cases nearly exact) solutions of the pairing problem using  realistic
hamiltonians - in particular those with the state-dependent pairing
hamiltonian. The new method is based on the direct solution of the many-body
problem in Fock space; it employs techniques similar to those used in the
nuclear shell-model, including the Lanchos diagonalisation scheme.  The
hamiltonian in question has the form:
\begin{equation}
       \hat{H}
       =
       \hat{H}_{mf}
       +
       \sum_{\alpha,\beta} 
       G_{\alpha,\beta} \;
       c^{\dagger}_{\alpha} c^{\dagger}_{\tilde\alpha}
       c^{\,}_{\tilde\beta} c^{\,}_{\beta} \; ,
                                                                  \label{eqn01}
\end{equation}
where $\hat{H}_{mf}$ denotes any mean-field hamiltonian e.g. the one with 
deformed Woods-Saxon potential and matrix $G_{\alpha,\beta}\,$, in general {\em
non-diagonal} in its indices, defines the state-dependent pairing-hamiltonian.
Owing to three exact symmetries obeyed by hamiltonians of the above general
form, cf. Ref.~\cite{mol} for details, the corresponding matrix written down
using a Fock space basis can be block-diagonalized analytically, thus reducing
the problem of diagonalisation of huge-size matrices to much smaller ones that
can be treated easily with the help of the Lanchos methods. 

In this article we report on the results of an introductory large-scale test
that consists in fitting the coupling constants of the simplest
(monopole-pairing) version of pairing hamiltonian in (\ref{eqn01}), i.e.
$G_{\alpha,\beta} = G\, \delta_{\alpha,\beta}$. Pairing calculations were
performed within Fock space defined by 24 particles placed on 24
double-degenerate single-particle energy levels in the 'pairing window'. The
corresponding full hamiltonian-matrix has the size of 
$N_{ham}=32\,247\,603\,683\,100$, while the sizes of hamiltonian-blocks, after
applying the formalism of Ref.~\cite{mol}, are: $N_{s=0}=2\,704\,156$ in the
seniority-zero and $N_{s=2}=705\,432$ in seniority-two blocks. In both cases we
have applied a basis cut-off reducing the sizes of the effectively diagonalized
matrices to $N^{\,\prime}_{s=0}=27\,703$ and  $N^{\,\prime}_{s=2}=26\,263$. We
have verified by comparison with the results of the exactly soluble method of
Richardson that the basis cut-off that reduces the matrix sizes by roughly two
orders of magnitude induces errors of $\sim 2 \%$ only.

\begin{figure}[ht]
       \epsfxsize=130mm \epsfbox{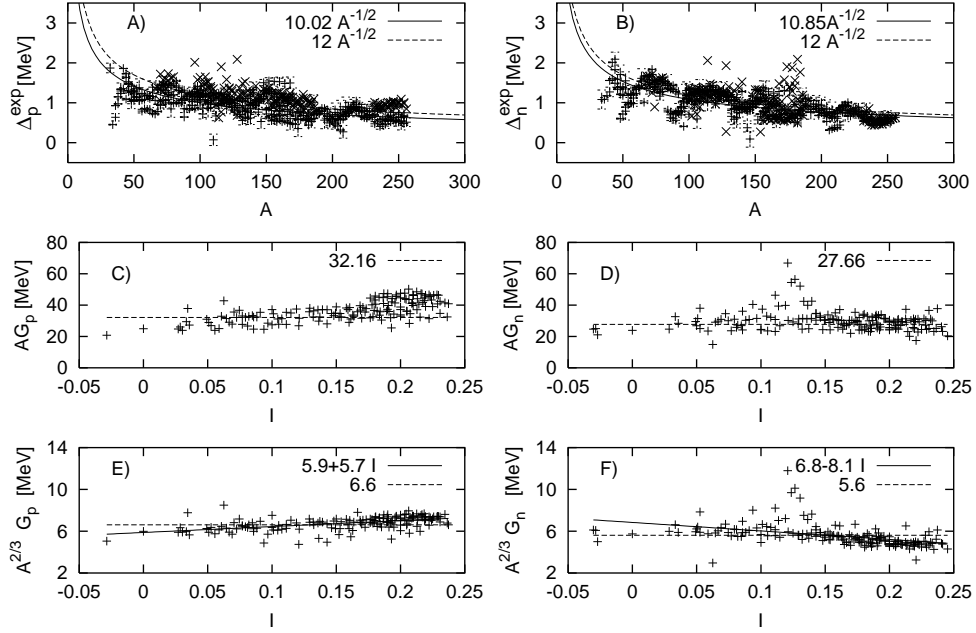}
       \caption{Experimental $\Delta^{(3)}_p$, protons, frame A), and  
                             $\Delta^{(3)}_n$, neutrons, frame B), together 
                with the average-fit curves of the form $\alpha/\sqrt{A}$ (top).
                The usually quoted average behavior $12/\sqrt{A}$ is also
                illustrated. Frames C) to F) represent the results of the 
                pairing constant-G fitting procedure (see the text and Table
                \ref{tab01}). Experimental masses are from Ref.~\cite{ant}.}
       \label{fig01}
\end{figure}

We have diagonalized the Fock-space hamiltonian-matrix corresponding to
(\ref{eqn01}) and we obtained the ground-state energy (seniority-zero) and the
lowest-energy seniority-two states. The difference between the two energies has
been interpreted as corresponding approximately to twice the 'pairing gap', the
latter obtained from the mass-difference expression
\begin{equation}
       \Delta_{n}^{(3)}(N) 
       = 
       {\pi_N\over 2} (B(N) + B(N + 2) - 2B(N+1))\,,
                                                                  \label{eqn02}
\end{equation}
as discussed recently in detail in Ref.~\cite{dob}. Above, $\pi_N = (-1)^N$, 
$B(N)$ are (negative) binding energies of nuclei, for the fixed $Z$-number.
To obtain the proton 'pairing gap' one has to replace $N$ with $Z$ and fix the
neutron number N. The resulting proton and neutron pairing gaps for even-even
nuclei are illustrated in Fig.~\ref{fig01}, top, in function of the mass number
$A$; the $\Delta^{(3)}$-values with the experimental errors exceeding 250 keV
were not taken into account. 

The best average fit in terms of the $\alpha/\sqrt{A}$ one-parameter dependence
for all the nuclei is $\Delta^{(3)}_n = 10.85/\sqrt{A}$ and $\Delta^{(3)}_p =
10.02/\sqrt{A}$ (solid lines in Fig.\ref{fig01}, top). The fitted
$\alpha$-values are smaller than those obtained in \cite{mad} from the liquid
drop model formula, the corresponding curves denoted in Fig.\ref{fig01} with
the dashed lines. One can also use alternative one-parameter expressions
similar to those in Ref.~\cite{hil}; the results of the fit are
\begin{equation}
      \Delta_n = 23/(\sqrt{A})^3\;   
      \quad {\rm and} \quad   
      \Delta_p = 21/(\sqrt{A})^3\; ,
                                                                  \label{eqn03}
\end{equation}

\vspace{-0.3truecm}\noindent
or

\vspace{-0.5truecm}
\begin{equation}
      \Delta_n =4.66/A^{1/3}    
      \quad {\rm and} \quad   
      \Delta_p =4.18/A^{1/3}\; ,
                                                                  \label{eqn04}
\end{equation}
but the fit precision remains similar to that with the $\alpha/\sqrt{A}$-type
dependence.

To avoid the undesired type of variation in pairing delta, characterististic
for spherical (especially doubly magic) nuclei,  the nuclei with available
experimental masses have been arbitrarily divided into 5 regions:  (I) with $Z
\in (32, 38)$ and $N \in (32, 44)$; (II) with $Z \in (40, 46)$ and $N \in (56,
72)$; (III) with $Z \in (54, 66)$ and $N \in (58, 76)$;  (IV) with $Z \in
(60,82)$ and $ N \in (88, 104)$ and (V) with $Z \in (90, 100)$ and $N \in (142,
156)$; there the doubly-magic nuclei have been eliminated. Within those regions
the $\Delta$-values were extracted and, as the next step, the $G$-constants
found that reproduce the extracted $\Delta$ values exactly for each of the
studied nuclei at the calculated in advance equilibrium deformations. The mean
field hamiltonian used is the same as in Ref.~\cite{wer} (see also references
there). The irregular behavior of $\Delta^{(3)}$ in function of $A$ suggests
that the resulting $G$ values will also vary in a relatively irregular fashion
and as a consequence we have tried 'a few parameter' fits in terms of $A$ and
$I \equiv (N-Z)/(N+Z)$ and possibly some powers of those variables:
\begin{equation}
       G_{p(n)} 
       = 
       \frac{\rho_{p(n)}}{A^s} (\rho_0 + \rho_1 I + \rho_2 I^2) \quad
       {\rm where} \quad
       s=1,1/2,2/3 \; .
                                                                  \label{eqn05}
\end{equation}
In Fig.~\ref{fig01}, frames C) and D), the products $G_{p(n)}\cdot A$ are shown
in function of the isospin factor I while the corresponding average behavior is
shown with the help of the solid lines. In frames E) and F) the quantities
$G_{p(n)} \cdot A^{2/3}$ are illustrated. The dependence in terms of $I$ 
obtained here is nearly constant. 
\begin{table}[ht]
\begin{center}
\small
\caption[T1b]{\label{tab01} Pairing strength constants in terms of the
              approximating expressions for neutrons and protons for 5 regions 
              considered.\\}             
\begin{tabular}{|c||r|r||r|r||r|r|r|}
\hline
$G_{p(n)}$ & \multicolumn{2}{c}{ $a/A + b I$} \vline& \multicolumn{2}{c}
                         { $(c+d\;I)/A^{2/3}$ } \vline& \multicolumn{3}{c}
                         { $(c^{\;\prime}+d^{\;\prime} I+eI^2)/A^{2/3}$} \vline\\
\hline
Region&$a$ &  $b$ &
$c$ &  $d$ & $c^{\prime}$ & $d^{\prime}$ & $e$   \\
\hline
$ all_{ n}$ & 24.872 &  0.147 &  6.517 &  -5.921 & 5.657 & 10.96 &  -64.18 \\
 I          & 24.230 & -0.001 &  5.887 &  -1.438 & 5.657 &  7.43 &  -65.63 \\
 II         & 27.177 & -0.165 &  6.230 &  -6.453 & 5.657 &  0.67 &  -21.39 \\
 III        & 33.129 & -0.214 &  7.097 & -10.298 & 5.657 & 19.60 & -152.21 \\
 IV         & 56.170 & -0.849 & 11.457 & -33.690 & 5.657 & 36.46 & -206.93 \\
 V          & 38.548 & -0.144 &  6.315 &  -6.408 & 5.657 & -0.29 &  -13.35 \\
\hline
\hline
$ all_{p}$  & 26.861 &  0.331 &  7.100 &  -2.905 & 6.529 &  2.65 &  -5.51 \\
 I          & 24.548 &  0.018 &  5.934 &   4.493 & 6.529 & -9.13 & 103.58 \\
 II         & 45.893 & -0.750 &  7.249 &  -2.892 & 6.529 &  1.74 &  -3.91 \\
 III        & 39.296 & -0.481 &  7.178 &  -3.700 & 6.529 &  4.55 &  -7.60 \\
 IV         & 54.546 & -0.608 &  7.289 &  -3.327 & 6.529 &  1.88 &  -4.47 \\
 V          & 78.763 & -0.635 &  8.333 &  -4.289 & 6.529 &  6.18 &  -8.67 \\
\hline
\end{tabular}					     
\end{center}					     
\end{table}

Results in Table \ref{tab01} give the overall r.m.s. deviations that are rather
small. The variations obtained, based on the experimental data concerning a
broad range of nuclei, show more structure than the simple parametrisations
tested can take care of. In particular, strong variations in parameter $e$ from
one nuclear range to another, cf. column 8 in the Table, indicates that the
$I^2$ fluctuations are too rapid to allow deducing any systematic trends in
this  context. 

The parametrisations summarized in the Table are 'ready to use' in conjunction
with the Fock-space diagonalisation method of Ref.~\cite{mol}.

\end{document}